\documentclass{PoS}

\newcommand{\ord}{{\cal O}}

\def\kpn{K^+\rightarrow\pi^+\nu\bar\nu}

\def\klpn{K_{L}\rightarrow\pi^0\nu\bar\nu}

\newcommand{\tev}{\, {\rm TeV}}

\newcommand{\be}{\begin{equation}}
\newcommand{\ee}{\end{equation}}

\title{Patterns of Flavour Violation in the RSc Model, the LHT Model and 
Supersymmetric Flavour Models
}
\ShortTitle{Patterns of Flavour Violation}

\author{\speaker{Andrzej J. Buras}\\
Technical University Munich, Physics Department, D-85748 Garching, Germany,\\
 TUM Institute for Advanced Study, D-80333 M\"unchen, Germany \\
   E-mail: \email{aburas@ph.tum.de}}

\abstract{
We summarize the results on patterns of flavour violation in a 
Randall-Sundrum model with custodial protection (RSc) and compare them
with those identified in the Littlest Higgs Model with T--parity (LHT)
and in a number of SUSY Flavour Models.
While $K$ decays play in this presentation a
prominent role, the inclusion of $B$ physics and lepton flavour 
violation is crucial in the distinction between these three
popular extensions of the Standard Model (SM) by means of flavour
physics. 
}

\FullConference{2009 KAON International Conference KAON09,\\
		 June 09 - 12 2009\\
		 Tsukuba, Japan}

\begin{document}

\section{Introduction}
Flavour changing neutral current processes (FCNC) will without any 
doubt  provide
a clear distinction between various new physics (NP) scenarios beyond the
SM once the data on these processes will improve in the coming decade.
The goal of this writing is to illustrate this fact by summarising patterns
of flavour violation identified in 2006-2009 in my group at the TUM through
intensive studies of a Randall--Sundrum model with custodial protection 
(RSc) \cite{Blanke:2008zb,Blanke:2008yr,Albrecht:2009xr,Buras:2009ka}, the Littlest Higgs Model 
with T-Parity (LHT) 
\cite{Blanke:2006sb,Blanke:2006eb,Blanke:2006xr,Blanke:2007db,Blanke:2007wr,Blanke:2007ee,Bigi:2009df,Blanke:2009am}
 and of a number of
SUSY flavour models (SF) \cite{Altmannshofer:2009ne}. These three prominent directions beyond
the SM
contain new sources
of CP and flavour violation implying thereby in certain cases spectacular
deviations from the SM expectations and more generally from the patterns
of flavour violation characteristic for models with minimal flavour
violation (MFV).

In Section 2 we will briefly describe the RSc scenario and the results for
FCNC processes obtained by us. The presentation of the LHT model in Section
3 is very short as an update of our efforts in this model appeared recently
\cite{Blanke:2009am}.
The same statement applies to SF models discussed in Section 4 with a very
detailed analysis of these models presented in \cite{Altmannshofer:2009ne}. 
A brief comparison of
these three NP scenarios and an outlook in Section 5 close this mini-review.
Due to space limitations the list of references is incomplete and 
I apologize for it already here. I will improve on it in my EPS09 talk.

\section{FCNC Processes in the RSc Model}
\subsection{Express Review of the RS Framework \cite{Randall:1999ee}}
The story takes place in a 5D spacetime, where the extra dimension is 
compactified to the interval $0\le y\le L$ with a warped metric given by
\begin{equation}
ds^2=e^{-2ky}\eta_{\mu\nu}dx^\mu dx^\nu - dy^2
\end{equation}
and $\eta_{\mu\nu}$ being the usual 4D Minkowski metric. There are two
branes, the UV brane $(y=0)$ and the IR brane $(y=L)$ and the {\it bulk} 
between them. Energy scales are suppressed by the warp factor $\exp(-ky)$
providing a natural solution to the gauge hierarchy problem or equivalently 
to the vast disparity of the Planck and electroweak scales 
\cite{Randall:1999ee}.

In the setup considered by us

 {\bf 1.} Fermions (quarks and leptons) and gauge bosons, to be specified later
on, live in the bulk ~\,~\,~\,~\cite{Gherghetta:2000qt,Chang:1999nh,Grossman:1999ra},

 {\bf 2.} the Higgs boson is localised on the IR brane,

 {\bf 3.}  Kaluza-Klein (KK) excitations of fermions and gauge bosons 
live close
to the IR brane,

 {\bf 4.} the SM fermion (zero mode) shape functions depend 
exponentially on the {\it bulk
mass} parameters $c_i$ that are characteristic for a given fermion
\begin{equation}
f^{(0)}\sim e^{(1/2-c_i)ky}.
\end{equation} 
Consequently fermions with $c_i>1/2$ are localized towards the UV brane, 
while
the ones with $c_i<1/2$ towards the IR brane. As the Higgs lives on the
IR brane, fermions with $c_i>1/2$ have only a small overlap with the
Higgs and aquire masses much smaller than the Higgs vacuum expectation
value (these are basically all SM fermions except the top quark). On the
other hand for $c_t<1/2$ the large top quark mass is generated. Strictly 
speaking bulk parameters and shape functions for left- and right-handed
fermions are involved and the elements of 4D Yukawa matrices are given
in terms of the elements of the 5D Yukawa matrices as follows:
\begin{equation}
(Y^{4D}_{u,d})_{ij}= (Y^{5D}_{u,d})_{ij}f_i^Qf_j^{u,d}
\end{equation}
with $(Y^{5D}_{u,d})_{ij}$ supposed to be anarchic.
$f_i^Q$ and $f_j^{u,d}$ are 
the shape functions of left-handed and 
right-handed fermions, respectively, evaluated at the IR brane, 
where the Higgs is
placed. This set up allows to generate the hierarchical structure of 
quark masses and CKM parameters in a natural manner with 
$(Y^{5D}_{u,d})_{ij}$ and $c_i$ being of order unity 
\cite{Grossman:1999ra,Gherghetta:2000qt}.
 The latter
are typically in the range $0.3\le c_i\le 0.6$

{\bf 5.} The shape functions of gluons and photons are flat due to exact
$SU(3)\times U(1)_Q$ invariance. On the other hand the corresponding
profiles of the $W^\pm$ and $Z^0$ gauge bosons, while being flat before
the electroweak symmetry breaking (EWSB), are distorted by
$v^2/M_{KK}^2$ corrections near the IR brane after EWSB took place. 
$M_{KK}$ is the KK scale.
The 4D picture of this happening is the mixing of the heavy KK gauge
bosons with the $W^\pm$ and $Z^0$ in the process of EWSB modifying 
thereby the couplings of the final light gauge boson mass eigenstates that
are interpreted as the SM $W^\pm$ and $Z^0$.

{\bf 6.} Finally, the interactions between fermions and gauge bosons are
given by the overlaps of the corresponding shape functions.

\subsection{First Phenomenological Implications}
This general setup with the SM gauge symmetry in the bulk has the
following phenomenological implications:

{\bf A.} The fact that fermions are localized at different positions in the bulk
and the shape functions of $Z$ and $W^\pm$ gauge bosons are distorted
near the IR brane after EWSB and the KK gauge bosons are peaked toward
the IR brane implies non--universalities (in flavour) in the gauge interactions
of fermions and gauge bosons. After the transformation to fermionic 
mass eigenstates FCNC transitions mediated by $Z$ and the heavy neutral KK 
gauge bosons including KK gluons take place at tree level. Thus the 
ordinary GIM mechanism is broken already at the tree level as opposed to
the SM where it is broken first at the one-loop level. Fortunately, the 
gauge-fermion interactions in the RS framework exhibit also hierarchies
and consequently a natural suppression of FCNC processes
takes place. This suppression of FCNC transitions 
is known under the name of the RS-GIM
\cite{Huber:2003tu,Agashe:2004cp}.
It saves
the RS framework from a total disaster but as we will see below it is
insufficient in the case of certain observables to provide appropriate
suppression of FCNC transitions in the presence of the KK scales in the
reach of the LHC.

{\bf B.} The presence of FCNC transitions at the tree level is accompanied by the 
breakdown of the unitarity of the CKM matrix. Physically this breakdown
originates from the mixing of the SM gauge bosons with the heavy KK gauge
bosons on the one hand and from the mixing of the SM fermions with the KK 
fermions on the other hand.

{\bf C.} The FCNC tree level exchanges of KK gluons imply the presence of the
operators with the left-right (LR) Dirac structure that are very strongly
suppressed in the SM and usually neglected. In the case of $\Delta F=2$
transitions, that is particle-antiparticle mixings, the Wilson 
coefficients of these operators are strongly enhanced through 
renormalization group effects and moreover in the case of $K^0-\bar K^0$
mixing their hadronic matrix elements are chirally enhanced. Together
these effects amount relative to the contribution of the SM LL 
operator to an enhancement
by a factor of 140 at the amplitude level if the KK scale at which
this operator is generated is in the ballpark of 3~TeV.

{\bf D.} The presence of three $3\times3$ hermitian bulk matrices $c^Q$, $c^u$ and
$c^d$ in addition to the usual Yukawa couplings implies 27 new flavour
and CP-violating parameters: 18 real mixing angles and 9 complex phases
\cite{Agashe:2004cp}.
Thus the RS framework goes far beyond MFV.

{\bf E.} Finally, the modifications in the electroweak gauge sector imply
 KK scales  $M_{KK}\ge 2 $~TeV and $M_{KK}\ge 10$~TeV in order to be
consistent with the bounds on the $S$ and the $T$ parameter, respectively.
Also NP contributions to the $Zb_L\bar b_L$ coupling become problematic.

The problems with electroweak parameters just mentioned lead several
authors \cite{Agashe:2003zs,Csaki:2003zu,Agashe:2006at}
to invent protection mechanisms that as we will see turned
out to be useful also for suppressing FCNC processes. We will now 
discuss one particular model of this type.
\subsection{A RS Model with Custodial Protection (RSc)}
A more realistic class of RS models with the lowest KK excitations
in the reach of the LHC has the following general structure. The bulk
gauge symmetry is extended in order to obtain the usual custodial 
$SU(2)$ symmetry on the IR brane \cite{Agashe:2003zs,Csaki:2003zu}. 
Additionally, in order to avoid
problems with large NP contributions to the $Z b_L\bar b_L$ coupling, 
fermions are put into $SU(2)_L\leftrightarrow SU(2)_R$ 
symmetric representations \cite{Agashe:2006at}. 
Thus the bulk 
symmetry group is now
\begin{equation}
G_{\rm bulk}=SU(3)_c\times SU(2)_L\times SU(2)_R\times U(1)_X\times P_{LR}
\end{equation}
with $P_{LR}$ denoting the discrete symmetry interchanging $L$ and $R$.
$U(1)_X$ allows to assign the usual hypercharges to quarks and leptons.
$SU(2)_R\times U(1)_X\times P_{LR}$ is broken by boundary conditions on the UV 
brane  down to $U(1)_Y$ and $SU(2)_L\otimes SU(2)_R$ is broken on the IR 
brane down to the custodial $SU(2)_V$ when the neutral component of 
the Higgs develops a vacuum 
expectation value. This also breaks the SM gauge group $SU(2)_L\times U(1)_Y$
down to $U(1)_Q$.

In this new set up the $T$ parameter and the $Z b_L\bar b_L$ coupling are
protected from receiving NP contributions at the tree level up to
small $P_{LR}$ symmetry breaking effects due to the 
boundary conditions on the UV brane.
It has also been pointed out in 
\cite{Blanke:2008zb,Blanke:2008yr,Buras:2009ka} that this construction 
when extended to
three quark generations allows to protect certain flavour violating
couplings from receiving tree level contributions. These are 
$Z\bar d_L^i d_L^j$ and $Z\bar u^i_R u^j_R$. On the other hand
$Z\bar d_R^i d_R^j$, $Z\bar u^i_L u^j_L$ and $W^+\bar u^i_L d_L^j$
remain unprotected.

These are the general properties of the class of models in question. 
To be more concrete, fermion representations under the symmetry group
have to be chosen. The particular fermion assignement in the model 
 worked out by us in \cite{Albrecht:2009xr} has been motivated by the 
analyses of 
\cite{Cacciapaglia:2006gp,Contino:2006qr,Carena:2006bn,Carena:2007ua}. In particular the fermion $SO(4)$ 
representations considered by us can easily 
be embedded into complete $SO(5)$ multiplets used in 
\cite{Contino:2006qr,Carena:2006bn,Carena:2007ua} in the
context of models with gauge-Higgs unification. Thus
 the SM doublets in our model
belong to the representations (2,2) under $SU(2)_L\times SU(2)_R$,
$u^i_R$ to (1,1) and $d_R^i$ to $(1,3)$. Because of the $P_{LR}$ 
symmetry also (3,1) representations for each generation not containing
any of the SM particles have to be added. The implication of this
assignement is the presence of KK quarks with electric charges $\pm 5/3$
that could in principle be discovered at the LHC. The fermion content
of this model is explicitly given in \cite{Albrecht:2009xr}, where also 
 a complete set of 
Feynman rules has been worked out.

As far as the 
gauge boson sector is concerned, in addition to the SM gauge bosons
the lightest KK gauge bosons are the KK--gluons, KK-photon and the 
electroweak KK gauge bosons $W^\pm_H$, $W^{\prime\pm}$, $Z_H$ and $Z^\prime$,
all with masses $M_{KK}$ around $2-3\tev$.
\subsection{Patterns of Flavour Violation}
\subsubsection{Preliminaries}
The first rough estimates of NP contributions to $\Delta F=2$ processes 
in RS scenarios can be found in \cite{Burdman:2002gr,Agashe:2004cp}.
However, the first more sophisticated analysis of these processes
has been performed by Csaki, Falkowski and Weiler (CFW)
\cite{Csaki:2008zd}, who included in their analysis
 the contributions of the dangerous LR
operators generated by the tree level KK gluon exchanges   and used the
model independent bounds on the corresponding Wilson coefficients from the
UTfit collaboration \cite{Bona:2007vi}.

Assuming the $Y^{5D}$ couplings to be anarchic and $\ord(1)$ and describing the 
hierarchy of quark masses and weak mixings solely by the fermion shape
functions, that is by the geometry in the fifth dimension, CFW found that
the data on $\varepsilon_K$ imply a lower bound on $M_{KK}$ to be roughly 
$20$~TeV, totally out of the LHC reach.

In view of this situation a  RS-TUM team \cite{Blanke:2008zb} 
has been built with the goal to 
investigate the amount of fine tuning in $Y^{5D}$ couplings necessary to
achieve an agreement of the theory with the data on $\varepsilon_K$ for
the lowest values $M_{KK}\approx (2-3)\tev$ that are consistent with 
EWP tests. Three additional RS-TUM teams 
\cite{Albrecht:2009xr,Blanke:2008yr,Buras:2009ka}
were supposed to study the
details of a particular RS model and subsequently perform a number of
phenomenological analyses.

As already advertised before, the electroweak and flavour structure of the
model, including Feynman rules, has been worked out in 
\cite{Albrecht:2009xr}. The
phenomenological applications of these results have been made in three 
separate analyses that we will summarise briefly now.
\subsubsection{Particle-Antiparticle Mixing}
In view of the problems with $\varepsilon_K$ we have first attacked $\Delta
F=2$ processes in \cite{Blanke:2008zb}. The main advances made in this paper 
 w.r.t previous
analyses are
\begin{itemize}
\item
Full renormalization group analysis of $K^0-\bar K^0$, $B^0_d-\bar B^0_d$ and
$B^0_s-\bar B^0_s$ mixings at the NLO level by means of the master formulae
for
the Wilson coefficients of the full set of operators given in 
\cite{Buras:2001ra}.
\item
The inclusion of the contributions from all SM gauge bosons and their lowest
KK excitations. The previous analyses concentrated on KK gluon contributions.
\item
Simultaneous phenomenology of $\varepsilon_K$, $\Delta M_K$, $\Delta M_{s,d}$,
$S_{\psi K_S}$, $A_{\rm SL}^q$ and $\Delta\Gamma_q$. 
\item
Relation of the RSc flavour model to the Froggatt-Nielsen approach 
\cite{Froggatt:1978nt}
that allowed
to derive analytic formulae for masses and mixings. An independent analysis of
this type can be found in \cite{Casagrande:2008hr}. This group provided also 
another look at $\varepsilon_K$ \cite{Bauer:2008xb}.
\item
Calculation of the amount of fine tuning of $Y^{5D}$ couplings, using the 
measure of Barbieri and Giudice $\Delta_{BG}(\varepsilon_K)$ 
\cite{Barbieri:1987fn}, necessary to satisfy
the data on $\varepsilon_K$ with $M_{KK}\approx 2-3\tev$.
\end{itemize}

The main results of these efforts are:

{\bf 1.} Confirmation of the CFW analysis \cite{Csaki:2008zd} 
for anarchic $Y^{5D}$ couplings:
purely geometrical description of the quark masses and mixings and the
measured value of $\varepsilon_K$ require $M_{KK}\ge20\tev$.

{\bf 2.} Identification of the regions of parameter space with
only modest fine tuning of $Y^{5D}$, for which electroweak precision
constraints and all $\Delta F=2$ constraints, in particular coming from 
$\varepsilon_K$ and $\Delta M_K$, are satisfied and the quark masses and
mixings correctly reproduced with $M_{KK}\approx 2-3\tev$.

{\bf 3.} The pattern of NP contributions to $\Delta F=2$ processes
         turns out to be as follows:
 \begin{itemize}
\item
$\varepsilon_K$ and $\Delta M_K$ are dominated by KK gluon tree level
exchanges and the operator $Q_2^{\rm LR}$ and not the standard LL operator 
$Q^{\rm VLL}_1$, both given as follows
\begin{equation}
Q_2^{\rm LR}=(\bar s P_L d) (\bar s P_R d), \qquad
Q_1^{VLL}=(\bar s\gamma_\mu P_L d) (\bar s\gamma_\mu P_L d), 
\end{equation}
with $P_{R,L}=(1\pm\gamma_5)/2$.
\item
In the case of $\Delta M_{d,s}$, $S_{\psi K_S}$ and $S_{\psi\phi}$ 
and generally in the case of $\Delta B=2$ observables, the LL operator
 $Q^{\rm VLL}_1$
and $Q_2^{\rm LR}$ compete with each other and both $Z_H$ and KK gluon
contributions have to be taken into account.
\item
Most interestingly the CP asymmetry in $B_s$ decays, $S_{\psi\phi}$, 
can reach values as high as the central value 0.8 in the data from CDF and D0 
collaborations
to be compared with its SM value 0.04.
\end{itemize}
\subsubsection{Rare K and B Decays}         
For the allowed region of parameter space satisfying $\Delta F=2$ 
constraints and corresponding to a moderate fine tuning of $Y^{5D}$ 
characterised by $\Delta_{BG}(\varepsilon_K)\le 20$ we have 
\cite{Blanke:2008yr}
\begin{itemize}
\item
calculated the branching ratios for $\kpn$, $\klpn$, $K_L\to\pi^0l^+l^-$,
$K_L\to\mu^+\mu^-$, $B_{s,d}\to\mu^+\mu^-$, $B\to X_{s,d}\nu\bar\nu$,
\item
investigated correlations between various $\Delta F=1$ and $\Delta F=2$
processes.
\end{itemize}

 The main results of our analysis are given as follows:

{\bf 1.} The NP contributions to the processes listed above 
        are dominated by tree level $Z$ exchanges,
        but as the left-handed couplings are protected, this dominance
        is governed by the right-handed $Z$ couplings that are not 
        protected by the custodial symmetry.

{\bf 2.} The branching ratios for $\kpn$, $\klpn$, $K_L\to\pi^0l^+l^-$
         can be enhanced relative to the SM expectations up to factors of
         1.6, 2.5 and 1.4, respectively, when only moderate fine tuning 
in $\varepsilon_K$ is required. Otherwise the enhancements can be larger.
          $Br(\kpn)$ and $Br(\klpn)$ can
         be simultaneously enhanced but this is not necessary as the
         correlations between these two branching ratios is not evident
         in this model. On the other hand $Br(\klpn)$ and 
         $Br(K_L\to\pi^0l^+l^-)$ ($l=e,\mu)$ are strongly correlated and the
         enhancement of one of these three branching ratios implies the
         enhancement of the remaining two.

{\bf 3.} A large enhancement of the short distance part of 
        $Br(K_L\to \mu^+\mu^-)$ is possible, up to a factor of 2-3, but
         not simultaneously with $Br(\kpn)$.

 {\bf 4.} More importantly simultaneous large NP effects in $S_{\psi\phi}$
          and $K\to\pi\nu\bar\nu$ channels are very unlikely.

{\bf 5.} The branching ratios for $B_{s,d}\to \mu^+\mu^-$ and 
             $B\to X_{s,d}\nu\bar\nu$ remain SM-like: the maximal enhancements
            of these branching ratios amount to $15\%$.

{\bf 6.} The relations \cite{Buras:2003jf,Buras:2009us}
         between various observables present in models with constrained 
        minimal violation can be strongly
         violated.
 \subsubsection{Impact of KK Fermions}
Until now our analysis did not include the contributions from KK fermions
that primarly affect the SM gauge-fermion couplings through their mixing 
with the SM fermions in the process of electroweak symmetry breaking.
In \cite{Buras:2009ka} using the effective Lagrangian approach and 
integrating out the KK
fermions from the low energy theory we have derived general
formulae for KK corrections to the SM couplings that involve quarks, $W^\pm$, 
$Z$ and the neutral Higgs. Our formulae can be applied to corrections from
 any vector-like fermions. Although they have been derived
in a different manner and have a different appearance, they turn out to be
equivalent to the ones presented earlier by del Aguila et al. 
\cite{delAguila:2000kb,delAguila:2000aa,delAguila:2000rc}.

Using these formulae we have demostrated explicitly that the custodial 
protection of $Z\bar d_L^i d_L^j$ and $Z\bar u_R^i u_R^j$ couplings 
remains valid in the presence of mixing with KK fermions which is guaranteed
by the $P_{LR}$ symmetric fermion representations.

Subsequently we have calculated the impact of KK fermions on the unprotected
couplings $Z\bar d_R^i d_R^j$, $Z\bar u_L^i u_L^j$, $W^+\bar u_L^i d_L^j$
and in particular analysed the violation of the unitarity of the CKM 
matrix and the generation of the right-handed couplings of the $W^\pm$
bosons $W^+\bar u_R^i d_R^j$ due to KK mixing.

Comparing these effects with the ones coming from KK gauge bosons analysed
earlier \cite{Blanke:2008zb,Blanke:2008yr}, we concluded that the 
latter effects are generally significantly
larger so that the impact of KK fermions with masses $\ord(M_{KK})$ on
our previous results is minor. In total, the corrections to the absolute
values of the
elements of the CKM matrix are very small except those involving the 
third generation quarks, where they can reach $1-2\%$. Also the corrections 
to the unitarity relations involving the third column and the third row 
are largest and can be as large as $5\%$.

Finally, we have verified that the effective Lagrangian approach in the
case at hand is rather accurate by diagonalizing numerically, in the
full theory, the relevant $18\times 18$, $12\times12$ and $18\times 12$
mass matrices.
\subsubsection{Summary}
The RS framework at large involves 18 new real flavour 
parameters and 9 new CP-violating phases in the quark sector that
reside in three $3\times3$ hermitian bulk matrices $c^Q$, $c^u$ and
$c^d$. It could appear then at first sight that this framework is not
predictive. Yet at least in the RSc model  a very 
clear pattern of flavour violation emerges from our studies after 
$\varepsilon_K$ in this model has been made consistent with the data
for $M_{KK}=2-3\tev$ with only moderate tuning of $Y^{5D}$:

{\bf A.} $S_{\psi\phi}$ can be much larger than in the SM and as
large as 0.8 in the ballpark of the central values found by CDF and
D0 collaborations.

{\bf B.} $Br(\klpn)$ and $Br(\kpn)$ can be enhanced up to factors 2.5
          and 1.6, respectively.

{\bf C.} Rare B decays turn out to be SM-like. In particular 
         $Br(B_{s,d}\to \mu^+\mu^-)$ can be enhanced by at most
         $15\%$.

{\bf D.} Simultaneous enhancements of $S_{\psi\phi}$ and of the 
         $K\to\pi\nu\bar\nu$ branching
         ratios are rather unlikely.

 This pattern implies that in the case of the confirmation of large
 values of $S_{\psi\phi}$ by future experiments  significant
 deviations of $Br(\klpn)$ and $Br(\kpn)$ from their SM values in
 this framework are very unlikely. 
On the other hand  SM-like value of $S_{\psi\phi}$
 will open the road for large enhancements of these branching ratios that
 could be tested by KOTO at J-Parc and NA62 at CERN, respectively. 
 Reviews of our work on the RSc model appeared in 
\cite{Duling:2009sf,Gori:2009tr,Buras:2009us,Blanke:2009mn,Duling:2009vc,Gori:2009em}. 

 Finally, let me just mention  that large NP contributions in the RS framework
 that require some tunings of parameters
in order to be in agreement with the experimental data have been found in 
$Br(B\rightarrow X_s\gamma)$ \cite{Agashe:2008uz}, $Br(\mu\rightarrow
 e\gamma)$ \cite{Agashe:2006iy,Davidson:2007si,Agashe:2009tu} and EDM's
 \cite{Agashe:2004cp,Iltan:2007sc}, that are all dominated by dipole operators. Also the new 
contributions to $\epsilon^{\prime}/{\epsilon}$ can
be large \cite{Gedalia:2009ws}.

New theoretical ideas addressing the issue of large FCNC transitions in the
RS framework and proposing new protection mechanism leading occasionally 
to MFV can be found in 
\cite{Csaki:2008eh,Cacciapaglia:2007fw,Cheung:2007bu,Santiago:2008vq,Csaki:2009bb,Csaki:2009wc}.

\section{LHT News}
\subsection{Preliminaries}
Other popular extensions of the SM is the Littlest Higgs model
without \cite{Arkani-Hamed:2002qy} and with T-parity 
\cite{Cheng:2003ju,Cheng:2004yc}  
in which the Higgs boson is protected by a new global symmetry,
possibly originating in a new fermionic system with new strong interactions
with the corresponding scale $\Lambda=\ord(10)\tev$. The SM Higgs boson is a
pseudo-Goldstone boson of this symmetry. In order to make this model 
consistent with electroweak precision tests and simultaneously having
the new particles of this model in the reach of the LHC, a discrete symmetry,
T-parity, has been introduced \cite{Cheng:2003ju,Cheng:2004yc}. 
Under T-parity all SM particles are {\it even}.
Among the new particles only a heavy $+2/3$ charged T quark belongs to the
even sector. Its role is to cancel the quadratic divergence in the Higgs
mass generated by the ordinary top quark. The even sector and also the 
model without T-parity do not go
beyond MFV \cite{Buras:2004kq,Buras:2006wk}.

More interesting from the point of view of FCNC processes is the T-odd
sector. It contains three doublets of mirror quarks and three doublets
of mirror leptons that communicate with the SM fermions by means of heavy 
$W^\pm_H$, $Z_H^0$ and $A^0_H$ gauge bosons. These interactions are 
governed by new mixing matrices $V_{Hd}$ and $V_{Hl}$ for down-quarks and
charged leptons, respectively. The corresponding matrices in the up ($V_{Hu}$)
and neutrino $(V_{H\nu})$ sectors are obtained by means of the relations
$V_{Hu}^\dagger V_{Hd}=V_{CKM}$, $V_{H\nu}^\dagger V_{Hl}=V^\dagger_{PMNS}$
\cite{Hubisz:2005bd,Blanke:2006xr}.

The following properties distinguish primarily this model from the RS
framework:

{\bf 1.} The model has a smaller number of new flavour parameters: 7 real 
         parameters and
         3 CP-violating phases in addition to 9+1 in the SM quark sector.

{\bf 2.} Only SM operators are relevant in this model so that the absence
         of dangerous LR operators allows to satisfy $\Delta F=2$ constraints
       basically without any significant fine tuning \cite{Blanke:2009am}. 
         Moreover non-perturbative
         uncertainties in this model are smaller than in the RSc framework as
        no new hadronic matrix elements are involved.

{\bf 3.} The NP scales are by a factor of $3-5$ lower than in the case of the
         RSc model, so that new fermions and new gauge bosons with masses
          below $1\tev$ are easily in the reach of the LHC.

{\bf 4.} There are no tree level FCNC transitions mediating $\Delta S=2$
         and $\Delta B=2$ transitions as well as rare $K$ and $B_{s,d}$
         decays. In the LHT model in all these processes
         the GIM mechanism is broken first at the one-loop level.

In 2006--2009 several LHT-TUM teams performed extensive analyzes of FCNC
processes in the quark and the lepton sector in the LHT model. Selected
reviews of our work can be found in 
\cite{Blanke:2007ww,Buras:2007zt,Duling:2007sf,Recksiegel:2009vj}.
Earlier we have also 
analysed the 
Littlest Higgs model without T-parity (LH) \cite{Buras:2004kq,Buras:2006wk}. 
This model has no custodial
protection relevant for EWP observables and the NP scale is shifted to
$2-3\tev$. These higher NP scales and the fact that
the LH model is of MFV type, makes this model phenomenologically less
interesting than the LHT model and we will not discuss it here.
\subsection{LHT Update}
One of the disturbing features of our 2006 results in the LHT model was the 
presence of an UV logarithmic divergence in the Z-penguin diagram that
we interpreted as the sensitivity to the unknown UV completion of this
model. Such an UV sensitivity should not really be surprising in models
of this type, in which the high energy theory is not specified, and in fact
has been identified first in the LH model \cite{Buras:2006wk}.

However it turns out that in deriving vertices in the
T-odd sector that contained right-handed mirror fermions and Z-boson we
have missed some $\ord(v^2/f^2)$ corrections identified in 2008 by Goto et al
\cite{Goto:2008fj} and del Aguila et al \cite{delAguila:2008zu} 
during their study 
of $K\to\pi\nu\bar\nu$ and
lepton flavour violating processes, respectively. The inclusion of this new
contribution cancels the divergences mentioned above making the LHT model 
much less sensitive to a possible UV completion of this model. While it has 
not been understood whether the disappearance of all UV divergences is
connected with the presence of T-parity, this is clearly a good news.

Recently our LHT-Rescue-Team \cite{Blanke:2009am} 
repeated all our previous phenomenological 
analyses, this time adding the missed corrections and updating the input
parameters. It should be emphasised that all our calculations not involving 
Z penguin diagrams like particle-antiparticle mixings, $b\to s\gamma$, 
$\mu\to e\gamma$, $\tau\to\mu\gamma$ and several other processes are not
affected by these findings. On the other hand the NP effects in
$\kpn$, $\klpn$, $K_L\to\pi^0l^+l^-$, $\mu\to 3e$, $\tau\to 3\mu$ and few
other transitions are now smaller than reported by us previously, although
they are still sizable.

The main results of our 2006--2009 efforts in the LHT model to be compared
with the outcome of our efforts in the RSc model summarized at the end of
Section 2 are as follows.

{\bf 1.} $S_{\psi\phi}$ can be much larger than its SM value but typically
 smaller than found in the RSc model: values above 0.3 are rather unlikely.

{\bf 2.} $Br(\klpn)$ and $Br(\kpn)$ can be enhanced up to factors of
 3 and 2.5, respectively. Therefore they can reach visibly larger values 
 than in the RSc model. The allowed points in the $Br(\klpn)$ vs $Br(\kpn)$
 plot cluster around two branches. On one of them $Br(\kpn)$ can reach 
 maximal values while $Br(\klpn)$ is SM-like. Here $Br(\kpn)$ can easily 
reach the central experimental value of E949 Collaboration at Brookhaven 
  \cite{Artamonov:2008qb}.
On the other one $Br(\klpn)$ 
 can reach maximal values but $Br(\kpn)$ can be enhanced by at most  
 a factor of 1.4 and therefore not reaching the central experimental 
 values. No such correlations are
found in the case of the RSc model and moreover in this model the 
maximal enhancements are smaller: 2.5 and 1.6 for $Br(\klpn)$ and 
$Br(\kpn)$, respectively. The central experimental value for the 
latter branching ratio therefore cannot be reached in this model.
Some insights in this different behaviour have recently been provided in
\cite{Blanke:2009pq}.

{\bf 3.} Rare B-decays turn out to be SM-like but they can deviate by more
from the SM expectations than it is the case for RSc. 
In particular $Br(B_{s,d}\to\mu^+\mu^-)$ can be enhanced by $30\%$ and a
significant part of this enhancement comes from the T-even sector.

{\bf 4.} Simultaneous enhancements of $S_{\psi\phi}$ and of
$Br(K\to\pi\nu\bar\nu)$ is similarly to the RSc scenario rather unlikely
but this feature is less pronounced than in the RSc model.

{\bf 5.} $Br(\mu\to e\gamma)$ in the LHT model can reach the upper bound
of $2\cdot 10^{-11}$ from the MEGA collaboration and in fact some fine
tuning of the parameters is required to satisfy this bound   
\cite{Blanke:2007db,delAguila:2008zu,Blanke:2009am}:
either the
corresponding mixing matrix in the mirror lepton sector has to be at
least as hierarchical as the CKM matrix and/or the masses of mirror
leptons carrying the same electric charge must be quasi-degenerate.
Therefore if the MEG collaboration does not find anything at the level of
$10^{-13}$, significant fine tuning of the LHT parameters 
will be required in order to keep $\mu\to e\gamma$ under control. Similar
comments apply to the RSc scenario.

{\bf 6.} It is not possible to distinguish the LHT model from the RSc and 
         the supersymmetric models discussed in Section 4 on the basis of 
         $\mu\to e\gamma$ alone. On the other hand as pointed out in 
\cite{Blanke:2007db} such a distinction can be made by measuring any of the 
           ratios $Br(\mu\to 3e)/Br(\mu\to e\gamma)$, 
          $Br(\tau\to 3\mu)/Br(\tau\to \mu\gamma)$, etc. In supersymmetric
          models all these decays are governed by dipole operators so 
         that these ratios are $\ord(\alpha)$ 
\cite{Ellis:2002fe,Arganda:2005ji,Brignole:2004ah,Paradisi:2005tk,Paradisi:2006jp}.
        In the
        LHT model the LFV decays with three leptons in the final state are
        not governed by dipole operators but by Z-penguins and box diagrams
        and the ratios in question turn out to be by at least one order of
        magnitude
        larger than in supersymmetric models.

 {\bf 7.} Recently also CP violation in  $D^0-\bar D^0$ mixing has 
been analysed in
the LHT model \cite{Bigi:2009df}. Observable effects at a level well 
beyond anything
possible with CKM dynamics have been identified. Comparisons with CP violation
in $K$ and $B$ systems should offer an excellent test of this NP scenario and
reveal the specific pattern of flavour and CP violation in the $D^0-\bar D^0$
system predicted by this model.

\section{Supersymmetric Flavour (SF) Models}
\subsection{Preliminaries}
The general MSSM framework with very many new flavour parameters in the
soft sector is not predictive and is plagued by flavour and CP problems:
FCNC processes and electric dipole moments are generically well above
the experimental data and upper bounds, respectively. Moreover the MSSM
framework addressing primarily the gauge hierarchy problem and the
quadratic divergences in the Higgs mass does not provide automatically the
hierarchical pattern of quark and lepton masses and of FCNC and CP 
violating interactions.

Much more interesting from this point of view are supersymmetric flavour
models with flavour symmetries that allow for a simultaneous understanding of 
the flavour structures in the Yukawa couplings and in SUSY soft-breaking 
terms, adequately suppressing FCNC and CP violating phenomena and solving
SUSY flavour and CP problems.

Supersymmetric models with flavour symmetries can be divided into two broad
classes depending on whether they are based on abelian or non-abelian 
flavour symmetries. Moreover, their phenomenological output crucially
depends on whether the flavour and CP violations are governed by left-handed
currents or there is an important new right-handed current component
 \cite{Altmannshofer:2009ne}.
They can be considered as generalisations of the Froggatt-Nielsen mechanism
for generating hierarchies in fermion masses and their interactions but
are phenomenologically much more successful than the original 
Froggatt-Nielsen model \cite{Froggatt:1978nt}. There is a rich 
literature on supersymmetric
models based on flavour symmetries and I do not have space to refer
to all of them here. A rather complete list of references can be 
found in a very recent paper from my group \cite{Altmannshofer:2009ne}  that I
will briefly summarise now. See also \cite{Altmannshofer:2009ap}.
\subsection{Patterns of Flavour Violation in the SF Models}
Recently one of the SUSY-TUM teams \cite{Altmannshofer:2009ne}
performed an extensive study of processes governed by $b\to s$ transitions 
in the SF models and of their correlations with processes governed by 
$b\to d$ transitions, 
$s\to d$ transitions, $D^0-\bar D^0$ oscillations, LFV 
decays, electric dipole moments and $(g-2)_{\mu}$. 
Both abelian and non-abelian flavour models have been considered as well as the
flavour blind MSSM (FBMSSM) and the MSSM with MFV. It has been shown how
 the characteristic patterns of correlations among the considered flavour 
observables allow to distinguish between these different SUSY scenarios and 
also to distinguish them from RSc and LHT scenarios of NP.

Of particular importance in our study were the correlations between 
the CP asymmetry $S_{\psi\phi}$ and
$B_s\rightarrow\mu^+\mu^-$, between the observed anomalies in 
$S_{\phi K_s}$ and $S_{\psi\phi}$, between 
$S_{\phi K_s}$ and $d_e$, between $S_{\psi\phi}$ and $(g-2)_{\mu}$ and 
also those involving LFV decays.

In the context of our study of the SF models we have analysed the 
following representative scenarios:
\begin{itemize}
\item [i)] dominance of right-handed (RH) currents 
(abelian model by Agashe and Carone\cite{Agashe:2003rj}),
\item [ii)] comparable left- and right-handed currents with CKM-like mixing
  angles represented by the special version (RVV2) 
of the non abelian $SU(3)$ 
model by
Ross, Velasco and Vives \cite{Ross:2004qn} as discussed recently in \cite{Calibbi:2009ja} and 
the model by Antusch, King and Malinsky (AKM) \cite{Antusch:2007re},
\item [iii)] dominance of left-handed (LH) currents in non-abelian 
models~\cite{Hall:1995es}.
\end{itemize}

In the choice of these three classes of flavour models, we were guided by our
model independent analysis, that I cannot discuss here because of the lack of 
space. Indeed  these three scenarios
predicting  quite 
different patterns of flavour violation should give a good representation of
most SF models discussed in the literature.
The distinct patterns of flavour violation found in each scenario have
been illustrated with several  plots that can  be found
in figures 11-14 of   \cite{Altmannshofer:2009ne}.

The main messages from our analysis of the models in question are as 
follows:

{\bf 1.}
Supersymmetric models with RH currents (AC, RVV2, AKM) and those with 
exclusively LH currents
can be globally distinguished by the values of the CP-asymmetries 
$S_{\psi\phi}$ and $S_{\phi K_S}$ with the following important result: 
none of
the models considered by us can simultaneously explain the $S_{\psi\phi}$ and
$S_{\phi K_S}$ anomalies observed in the data. 
In the models with RH currents,
$S_{\psi\phi}$ can naturally be much larger than its SM value, while 
$S_{\phi K_S}$ remains either SM-like or its correlation with $S_{\psi\phi}$ 
is inconsistent with the data. 
On the contrary, in the models with LH currents only,
$S_{\psi\phi}$ remains SM-like, while the  $S_{\phi K_S}$  anomaly can be 
easily explained.
Thus already future precise measurements of 
$S_{\psi\phi}$ and $S_{\phi K_S}$ will select one of these two classes of 
models, if any.

{\bf 2.}
The desire to explain the $S_{\psi\phi}$ anomaly within the models with
RH currents unambiguously implies, in the case of the AC and the AKM models,
values of
$Br(B_s\to\mu^+\mu^-)$ as high as several $10^{-8}$. In the 
RVV2 model such values are also possible but not necessarily implied
by the large value of $S_{\psi\phi}$. However, in all these models large
values of $S_{\psi\phi}$ imply automatically the 
solution to the $(g-2)_\mu$ anomaly. Moreover, the ratio 
$Br(B_d\to\mu^+\mu^-)/Br(B_s\to\mu^+\mu^-)$ in the AC and RVV2  models is 
dominantly below its MFV prediction and can be much smaller than the latter.
In the AKM model this ratio stays much closer to the MFV value of roughly 
$1/33$ \cite{Buras:2003td,Hurth:2008jc} and can be smaller or larger than 
this value with equal probability.
Still, values of $Br(B_d\to\mu^+\mu^-)$ as high as $1\times 10^{-9}$ are
possible in all these models.

{\bf 3.}
In the RVV2 and the AKM models, a large value of $S_{\psi\phi}$  combined with
the desire to explain the $(g-2)_\mu$ anomaly implies 
$Br(\mu\to e\gamma)$ in 
the reach of the MEG experiment.  In the case of the RVV2 model,
$d_e\ge 10^{-29}$ e.cm. is predicted, while in the AKM model it is typically
smaller.
 Moreover, in the case of the RVV2 model, 
$Br(\tau\to\mu\gamma)\ge 10^{-9}$ is then
 in the reach of Super-B machines, while this is not the case in the AKM model.

{\bf 4.}
Next, while the abelian AC model resolves the present UT tensions 
\cite{Lunghi:2008aa,Buras:2008nn,Lunghi:2009sm,Buras:2009pj}
through the modification of the ratio $\Delta M_d/\Delta M_s$, the 
non-abelian flavour models RVV2 and AKM provide the solution through
NP contributions to $\epsilon_K$. Moreover, while the AC model predicts
sizable NP contributions to $D^0-\bar D^0$ mixing, such contributions 
are tiny in the RVV2 and AKM models.

{\bf 5.} The hadronic EDMs represent very sensitive probes of SUSY flavour 
models with right-handed
currents. In the AC model, large values for the neutron EDM might be easily 
generated by both the
up- and strange-quark (C)EDM. In the former case, visible CPV effects in 
$D^0-\bar D^0$ mixing
are also expected while in the latter case large CPV effects in the
$B_s$ system are unavoidable.
The RVV2 and AKM models predict values for the down-quark (C)EDM and, hence for the neutron
EDM, above the $\approx 10^{-28}e$~cm level 
when a large $S_{\psi\phi}$
is generated. All the above models predict a large strange-quark (C)EDM, hence, a reliable
knowledge of its contribution to the hadronic EDMs, by means of lattice QCD techniques, would
be of the utmost importance to probe or to falsify flavour models embedded in a SUSY framework.

{\bf 6.}
In the supersymmetric models with exclusively LH currents, 
the desire to explain
the $S_{\phi K_S}$ anomaly implies automatically the solution to the 
$(g-2)_\mu$ anomaly and the direct CP asymmetry in $b\to s\gamma$ much
larger than its SM value. This is in contrast to the models with RH currents
where this asymmetry remains SM-like.

{\bf 7.}
Interestingly,  in the LH-current-models, the ratio 
$Br(B_d\to\mu^+\mu^-)/Br(B_s\to\mu^+\mu^-)$ can not only deviate
significantly from its MFV value of approximately $1/33$, 
but in contrast to the models with 
RH currents considered by us can also be much larger than the latter value. 
Consequently,
$Br(B_d\to\mu^+\mu^-)$ as high as $(1-2)\times 10^{-9}$ is still 
possible while being consistent with the bounds on all other observables,
in particular the one on $Br(B_s\to\mu^+\mu^-)$. Also interesting
correlations between $S_{\phi K_S}$ and CP asymmetries in 
$B\to K^*\ell^+\ell^-$ are found.

{\bf 8.}
Finally the branching ratios for 
$K\to\pi \nu\bar\nu$ decays in the supersymmetric models considered by us
remain SM-like and can be distinguished from RSc and LHT models where  
they can be significantly enhanced.

\section{Summary and Outlook}
In our search for a fundamental theory the understanding of the observed
flavour and CP  violating interactions and the explanation of the 
hierarchical fermion spectrum is a very important goal. In this mini-review
we have addressed in some detail FCNC processes in the RSc model, briefly
summarized TUM results for these processes in the LHT model and made a
propaganda for our recent very detailed analysis of FCNC processes, 
electric dipole moments and $(g-2)_\mu$ in a number of supersymmetric
flavour models.

The patterns of flavour and CP violation in NP scenarios considered 
by us are in spite of many parameters involved sufficiently distinct 
that they can be distinguished from each other through experiments in
the coming decade. This assumes that significant departures from the SM
expectations will be found in a number of observables in $K$, $B_{s,d}$ 
and $D$ systems and in particular in LFV decays and EDMs. Also the 
persistent $(g-2)_\mu$ anomaly will play an important role in these
considerations.

As an overture to various future possibilities let us consider the following
two scenarios which could turn out soon to be  reality.

{\bf Scenario 1.} Let us assume that $S_{\psi\phi}$ has been found to 
be $0.30\pm 0.05$, 
well beyond anything achievable by the CKM dynamics, but naturally 
found in the RSc model, in two SF models with RH currents (AC,RVV2) and
with some efforts in the AKM and the LHT model.
The SF models with LH-currents only, having $S_{\psi\phi}$ SM-like, 
will be ruled out.

First, this is rather bad news for the $K\to\pi\nu\bar\nu$ experiments
if any of these models turns out to be chosen by nature:
$Br(\klpn)$ and $Br(\kpn)$ will be SM-like in all these models.
In the RSc and LHT models this is a consequence of a large $S_{\psi\phi}$.
In the SF models $K\to\pi\nu\bar\nu$ decays are SM--like as the relevant NP
effects are strongly suppressed in these models in the process of solving
the SUSY-FCNC problem. Yet, there are other supersymmetric models
where these decays can be strongly enhanced. But this story is for another
occassion.

Now the distinction between RSc, LHT, AC, RVV2 and AKM models will be partly
made through the measurement of $Br(B_s\to \mu^+\mu^-)$. If this branching
ratio will turn out to be $\ord (10^{-8})$, RSc and LHT will be ruled out,
while such an enhancement of  $Br(B_s\to \mu^+\mu^-)$ is a prediction of
the AC and AKM models. In the RVV2 model this is not a prediction but
such high values can be accomodated in this model. The AC and AKM models 
can then be distinguished through the manner they address various tensions
in the standard UT analyses. While the AC model would favour 
$\gamma > 80^\circ$ and $\alpha < 80^\circ$ without NP contributions
to $\varepsilon_K$, the AKM model can solve the recent $\varepsilon_K$ 
anomaly \cite{Buras:2008nn,Buras:2009pj}
 with NP contributions to $\varepsilon_K$, while keeping 
$\gamma\approx 70^\circ$ and $\alpha \approx 90^\circ$ as suggested by 
UTfits and CKMfitter. Thus precise measurements of the angle $\gamma$ 
at the LHCb could select one of these models. Another distinction could 
be made through CP violation in the $D^0-\bar D^0$ oscillations that
is predicted to be sizable in the AC model but small in the AKM model
\cite{Altmannshofer:2009ne}.

{\bf Scenario 2.} Let us next assume that $S_{\psi\phi}$ has been found to 
be $0.30\pm 0.05$, as in the previous example, but $Br(B_s\to \mu^+\mu^-)$
turns out to be SM--like, that is in the ballpark of $4\cdot 10^{-9}$.
AC and AKM models are then ruled out but RSc and LHT models will be
in a good shape as they both expect $Br(B_s\to \mu^+\mu^-)$ to be 
SM-like. Similarly the RVV2 model will survive as it can accomodate
such low values of this branching ratio in the presence of a substantial
$S_{\psi\phi}$. The distinction between these three models can then be
made by means of other observables but the upper bound on the number
of pages for this contribution set by the organizers of KAON 09 has already
been badly violated and I will not discuss it here. 
Readers, who are are
still reading this sentence, are asked 
to look into our analysis in \cite{Altmannshofer:2009ne}, where 
a DNA-Flavour Test has been proposed.
This test should give still a deeper insight into
the patterns of flavour violation in various scenarios, in particular when
considered simultaneously  with various correlations present in concrete
models. The interplay of these efforts with the direct searches for NP will 
be most exciting.

Let me then close this mini-review of our work at TUM with the following
note. In spite of no spectacular discoveries of NP in flavour violating
processes this year, it was a very interesting workshop, very well 
organized and kept in a very pleasent atmosphere. I am happy I could be
a part of it. While thanking the organisers of KAON 09 for inviting me
to give this talk, I would also like to thank the members of various 
NP-TUM teams for very fruitful collaborations without which this talk would
not be possible. Special thanks go to Monika Blanke for critical comments 
on the manuscript.

This research was partially supported by the Deutsche
Forschungsgemeinschaft (DFG) under contract BU 706/2-1, the DFG Cluster of
Excellence `Origin and Structure of the Universe' and by the German
Bundesministerium f{\"u}r Bildung und Forschung under contract 05HT6WOA.


\begin{thebibliography}{99}

\bibitem{Blanke:2008zb}
M.~Blanke, A.~J. Buras, B.~Duling, S.~Gori, and A.~Weiler, {\it {$\Delta$ F=2
  Observables and Fine-Tuning in a Warped Extra Dimension with Custodial
  Protection}},  {\em JHEP} {\bf 03} (2009) 001,
  [\href{http://xxx.lanl.gov/abs/0809.1073}{{\tt arXiv:0809.1073}}].


\bibitem{Blanke:2008yr}
M.~Blanke, A.~J. Buras, B.~Duling, K.~Gemmler, and S.~Gori, {\it {Rare K and B
  Decays in a Warped Extra Dimension with Custodial Protection}},  {\em JHEP}
  {\bf 03} (2009) 108, [\href{http://xxx.lanl.gov/abs/0812.3803}{{\tt
  arXiv:0812.3803}}].

\bibitem{Albrecht:2009xr}
M.~E. Albrecht, M.~Blanke, A.~J. Buras, B.~Duling, and K.~Gemmler, {\it
  {Electroweak and Flavour Structure of a Warped Extra Dimension with Custodial
  Protection}},  \href{http://xxx.lanl.gov/abs/0903.2415}{{\tt
  arXiv:0903.2415}}.

\bibitem{Buras:2009ka}
A.~J. Buras, B.~Duling, and S.~Gori, {\it {The Impact of Kaluza-Klein Fermions
  on Standard Model Fermion Couplings in a RS Model with Custodial
  Protection}},  \href{http://xxx.lanl.gov/abs/0905.2318}{{\tt
  arXiv:0905.2318}}.

\bibitem{Blanke:2006sb}
M.~Blanke {\em et.~al.}, {\it {Particle antiparticle mixing, $\epsilon_K$,
  $\Delta\Gamma_q$, $A_{\rm SL}^q$, $A_{\rm CP}(B_d\to \psi K_S)$, $A_{\rm
  CP}(B_s\to \psi \phi)$ and $B\to X_{s,d} \gamma$ in the littlest Higgs model
  with T-parity}},  {\em JHEP} {\bf 12} (2006) 003,
  [\href{http://xxx.lanl.gov/abs/hep-ph/0605214}{{\tt hep-ph/0605214}}].

\bibitem{Blanke:2006eb}
M.~Blanke {\em et.~al.}, {\it {Rare and CP-violating K and B decays in the
  Littlest Higgs model with T-parity}},  {\em JHEP} {\bf 01} (2007) 066,
  [\href{http://xxx.lanl.gov/abs/hep-ph/0610298}{{\tt hep-ph/0610298}}].


\bibitem{Blanke:2006xr}
M.~Blanke {\em et.~al.}, {\it {Another Look at the Flavour Structure of the
  Littlest Higgs Model with T-Parity}},  {\em Phys. Lett.} {\bf B646} (2007)
  253--257, [\href{http://xxx.lanl.gov/abs/hep-ph/0609284}{{\tt
  hep-ph/0609284}}].

\bibitem{Blanke:2007db}
M.~Blanke, A.~J. Buras, B.~Duling, A.~Poschenrieder, and C.~Tarantino, {\it
  {Charged Lepton Flavour Violation and $(g-2)_\mu$ in the Littlest Higgs Model
  with T-Parity: a clear Distinction from Supersymmetry}},  {\em JHEP} {\bf 05}
  (2007) 013, [\href{http://xxx.lanl.gov/abs/hep-ph/0702136}{{\tt
  hep-ph/0702136}}].

\bibitem{Blanke:2007wr}
M.~Blanke, A.~J. Buras, S.~Recksiegel, C.~Tarantino, and S.~Uhlig, {\it
  {Correlations between $\epsilon^\prime$ / $\epsilon$ and rare $K$ decays in
  the littlest Higgs model with T--parity}},  {\em JHEP} {\bf 06} (2007) 082,
  [\href{http://xxx.lanl.gov/abs/0704.3329}{{\tt arXiv:0704.3329}}].

\bibitem{Blanke:2007ee}
M.~Blanke, A.~J. Buras, S.~Recksiegel, C.~Tarantino, and S.~Uhlig, {\it
  {Littlest Higgs Model with T--Parity Confronting the New Data on $D^0$ -
  $\bar{D}^0$ Mixing}},  {\em Phys. Lett.} {\bf B657} (2007) 81--86,
  [\href{http://xxx.lanl.gov/abs/hep-ph/0703254}{{\tt hep-ph/0703254}}].

\bibitem{Bigi:2009df}
I.~I. Bigi, M.~Blanke, A.~J. Buras, and S.~Recksiegel, {\it {CP Violation in
  $D^0 - \bar D^0$ Oscillations: General Considerations and Applications to the
  Littlest Higgs Model with T-Parity}},  {\em JHEP} {\bf 07} (2009) 097,
  [\href{http://xxx.lanl.gov/abs/0904.1545}{{\tt arXiv:0904.1545}}].


\bibitem{Blanke:2009am}
M.~Blanke, A.~J. Buras, B.~Duling, S.~Recksiegel, and C.~Tarantino, {\it {FCNC
  Processes in the Littlest Higgs Model with T-Parity: a 2009 Look}},
  \href{http://xxx.lanl.gov/abs/0906.5454}{{\tt arXiv:0906.5454}}.


\bibitem{Altmannshofer:2009ne}
W.~Altmannshofer, A.~J. Buras, S.~Gori, P.~Paradisi, and D.~M. Straub, {\it
  {Anatomy and Phenomenology of FCNC and CPV Effects in SUSY Theories}},
  \href{http://xxx.lanl.gov/abs/0909.1333}{{\tt arXiv:0909.1333}}.




\bibitem{Randall:1999ee}
L.~Randall and R.~Sundrum, {\it {A large mass hierarchy from a small extra
  dimension}},  {\em Phys. Rev. Lett.} {\bf 83} (1999) 3370--3373,
  [\href{http://xxx.lanl.gov/abs/hep-ph/9905221}{{\tt hep-ph/9905221}}].

\bibitem{Gherghetta:2000qt}
T.~Gherghetta and A.~Pomarol, {\it {Bulk fields and supersymmetry in a slice of
  AdS}},  {\em Nucl. Phys.} {\bf B586} (2000) 141--162,
  [\href{http://xxx.lanl.gov/abs/hep-ph/0003129}{{\tt hep-ph/0003129}}].

\bibitem{Chang:1999nh}
S.~Chang, J.~Hisano, H.~Nakano, N.~Okada, and M.~Yamaguchi, {\it {Bulk standard
  model in the Randall-Sundrum background}},  {\em Phys. Rev.} {\bf D62} (2000)
  084025, [\href{http://xxx.lanl.gov/abs/hep-ph/9912498}{{\tt
  hep-ph/9912498}}].

\bibitem{Grossman:1999ra}
Y.~Grossman and M.~Neubert, {\it {Neutrino masses and mixings in
  non-factorizable geometry}},  {\em Phys. Lett.} {\bf B474} (2000) 361--371,
  [\href{http://xxx.lanl.gov/abs/hep-ph/9912408}{{\tt hep-ph/9912408}}].

\bibitem{Huber:2003tu}
S.~J. Huber, {\it {Flavor violation and warped geometry}},  {\em Nucl. Phys.}
  {\bf B666} (2003) 269--288,
  [\href{http://xxx.lanl.gov/abs/hep-ph/0303183}{{\tt hep-ph/0303183}}].


\bibitem{Agashe:2004cp}
K.~Agashe, G.~Perez, and A.~Soni, {\it {Flavor structure of warped extra
  dimension models}},  {\em Phys. Rev.} {\bf D71} (2005) 016002,
  [\href{http://xxx.lanl.gov/abs/hep-ph/0408134}{{\tt hep-ph/0408134}}].

\bibitem{Agashe:2003zs}
K.~Agashe, A.~Delgado, M.~J. May, and R.~Sundrum, {\it {RS1, custodial isospin
  and precision tests}},  {\em JHEP} {\bf 08} (2003) 050,
  [\href{http://xxx.lanl.gov/abs/hep-ph/0308036}{{\tt hep-ph/0308036}}].

\bibitem{Csaki:2003zu}
C.~Csaki, C.~Grojean, L.~Pilo, and J.~Terning, {\it {Towards a realistic model
  of Higgsless electroweak symmetry breaking}},  {\em Phys. Rev. Lett.} {\bf
  92} (2004) 101802, [\href{http://xxx.lanl.gov/abs/hep-ph/0308038}{{\tt
  hep-ph/0308038}}].

\bibitem{Agashe:2006at}
K.~Agashe, R.~Contino, L.~Da~Rold, and A.~Pomarol, {\it {A custodial symmetry
  for $Z b \bar b$}},  {\em Phys. Lett.} {\bf B641} (2006) 62--66,
  [\href{http://xxx.lanl.gov/abs/hep-ph/0605341}{{\tt hep-ph/0605341}}].


\bibitem{Cacciapaglia:2006gp}
G.~Cacciapaglia, C.~Csaki, G.~Marandella, and J.~Terning, {\it {A New Custodian
  for a Realistic Higgsless Model}},  {\em Phys. Rev.} {\bf D75} (2007) 015003,
  [\href{http://xxx.lanl.gov/abs/hep-ph/0607146}{{\tt hep-ph/0607146}}].


\bibitem{Contino:2006qr}
R.~Contino, L.~Da~Rold, and A.~Pomarol, {\it {Light custodians in natural
  composite Higgs models}},  {\em Phys. Rev.} {\bf D75} (2007) 055014,
  [\href{http://xxx.lanl.gov/abs/hep-ph/0612048}{{\tt hep-ph/0612048}}].

\bibitem{Carena:2006bn}
M.~S. Carena, E.~Ponton, J.~Santiago, and C.~E.~M. Wagner, {\it {Light
  Kaluza-Klein states in Randall-Sundrum models with custodial SU(2)}},  {\em
  Nucl. Phys.} {\bf B759} (2006) 202--227,
  [\href{http://xxx.lanl.gov/abs/hep-ph/0607106}{{\tt hep-ph/0607106}}].



\bibitem{Carena:2007ua}
M.~S. Carena, E.~Ponton, J.~Santiago, and C.~E.~M. Wagner, {\it {Electroweak
  constraints on warped models with custodial symmetry}},  {\em Phys. Rev.}
  {\bf D76} (2007) 035006, [\href{http://xxx.lanl.gov/abs/hep-ph/0701055}{{\tt
  hep-ph/0701055}}].

\bibitem{Burdman:2002gr}
G.~Burdman, {\it {Constraints on the bulk standard model in the Randall-
  Sundrum scenario}},  {\em Phys. Rev.} {\bf D66} (2002) 076003,
  [\href{http://xxx.lanl.gov/abs/hep-ph/0205329}{{\tt hep-ph/0205329}}].

\bibitem{Csaki:2008zd}
C.~Csaki, A.~Falkowski, and A.~Weiler, {\it {The Flavor of the Composite
  Pseudo-Goldstone Higgs}},  {\em JHEP} {\bf 09} (2008) 008,
  [\href{http://xxx.lanl.gov/abs/0804.1954}{{\tt arXiv:0804.1954}}].

\bibitem{Bona:2007vi}
{\bf UTfit} Collaboration, M.~Bona {\em et.~al.}, {\it {Model-independent
  constraints on $\Delta$ F=2 operators and the scale of new physics}},  {\em
  JHEP} {\bf 03} (2008) 049, [\href{http://xxx.lanl.gov/abs/0707.0636}{{\tt
  arXiv:0707.0636}}].

\bibitem{Buras:2001ra}
A.~J. Buras, S.~Jager, and J.~Urban, {\it {Master formulae for $\Delta F = 2$
  NLO-QCD factors in the standard model and beyond}},  {\em Nucl. Phys.} {\bf
  B605} (2001) 600--624, [\href{http://xxx.lanl.gov/abs/hep-ph/0102316}{{\tt
  hep-ph/0102316}}].

\bibitem{Froggatt:1978nt}
C.~D. Froggatt and H.~B. Nielsen, {\it {Hierarchy of Quark Masses, Cabibbo
  Angles and CP Violation}},  {\em Nucl. Phys.} {\bf B147} (1979) 277.


\bibitem{Casagrande:2008hr}
S.~Casagrande, F.~Goertz, U.~Haisch, M.~Neubert, and T.~Pfoh, {\it {Flavor
  Physics in the Randall-Sundrum Model: I. Theoretical Setup and Electroweak
  Precision Tests}},  {\em JHEP} {\bf 10} (2008) 094,
  [\href{http://xxx.lanl.gov/abs/0807.4937}{{\tt arXiv:0807.4937}}].

\bibitem{Bauer:2008xb}
M.~Bauer, S.~Casagrande, L.~Gruender, U.~Haisch, and M.~Neubert, {\it {Little
  Randall-Sundrum models: $\varepsilon_K$ strikes again}},
  \href{http://xxx.lanl.gov/abs/0811.3678}{{\tt arXiv:0811.3678}}.

\bibitem{Barbieri:1987fn}
R.~Barbieri and G.~F. Giudice, {\it {Upper Bounds on Supersymmetric Particle
  Masses}},  {\em Nucl. Phys.} {\bf B306} (1988) 63.

\bibitem{Buras:2003jf}
A.~J. Buras, {\it {Minimal flavor violation}},  {\em Acta Phys. Polon.} {\bf
  B34} (2003) 5615--5668, [\href{http://xxx.lanl.gov/abs/hep-ph/0310208}{{\tt
  hep-ph/0310208}}].

\bibitem{Buras:2009us}
A.~J. Buras, {\it {Testing the CKM Picture of Flavour and CP Violation in Rare
  K and B Decays and Particle-Antiparticle Mixing}},
  \href{http://xxx.lanl.gov/abs/0904.4917}{{\tt arXiv:0904.4917}}.

\bibitem{delAguila:2000kb}
F.~del Aguila and J.~Santiago, {\it {Universality limits on bulk fermions}},
  {\em Phys. Lett.} {\bf B493} (2000) 175--181,
  [\href{http://xxx.lanl.gov/abs/hep-ph/0008143}{{\tt hep-ph/0008143}}].

\bibitem{delAguila:2000aa}
F.~del Aguila, M.~Perez-Victoria, and J.~Santiago, {\it {Effective description
  of quark mixing}},  {\em Phys. Lett.} {\bf B492} (2000) 98--106,
  [\href{http://xxx.lanl.gov/abs/hep-ph/0007160}{{\tt hep-ph/0007160}}].


\bibitem{delAguila:2000rc}
F.~del Aguila, M.~Perez-Victoria, and J.~Santiago, {\it {Observable
  contributions of new exotic quarks to quark mixing}},  {\em JHEP} {\bf 09}
  (2000) 011, [\href{http://xxx.lanl.gov/abs/hep-ph/0007316}{{\tt
  hep-ph/0007316}}].

\bibitem{Duling:2009sf}
B.~Duling, {\it {K and B meson mixing in warped extra dimensions with custodial
  protection}},  \href{http://xxx.lanl.gov/abs/0901.4599}{{\tt
  arXiv:0901.4599}}.

\bibitem{Gori:2009tr}
S.~Gori, {\it {Patterns of Flavour Violation in a Warped Extra Dimensional
  Model with Custodial Protection}},
  \href{http://xxx.lanl.gov/abs/0901.4704}{{\tt arXiv:0901.4704}}.

\bibitem{Blanke:2009mn}
M.~Blanke, {\it {K and B Physics in the Custodially Protected Randall- Sundrum
  Model}},  \href{http://xxx.lanl.gov/abs/0908.2716}{{\tt arXiv:0908.2716}}.

\bibitem{Duling:2009vc}
B.~Duling, {\it {Predictions for Flavour Observables in a RS Model with
  Custodial Symmetry}},  \href{http://xxx.lanl.gov/abs/0908.3099}{{\tt
  arXiv:0908.3099}}.


\bibitem{Gori:2009em}
S.~Gori, {\it {Patterns of Flavor Violation in a Warped Extra Dimensional Model
  with Custodial Protection}},  \href{http://xxx.lanl.gov/abs/0909.3042}{{\tt
  arXiv:0909.3042}}.


\bibitem{Agashe:2008uz}
K.~Agashe, A.~Azatov, and L.~Zhu, {\it {Flavor Violation Tests of
  Warped/Composite SM in the Two- Site Approach}},
  \href{http://xxx.lanl.gov/abs/0810.1016}{{\tt arXiv:0810.1016}}.


\bibitem{Agashe:2006iy}
K.~Agashe, A.~E. Blechman, and F.~Petriello, {\it {Probing the Randall-Sundrum
  geometric origin of flavor with lepton flavor violation}},  {\em Phys. Rev.}
  {\bf D74} (2006) 053011, [\href{http://xxx.lanl.gov/abs/hep-ph/0606021}{{\tt
  hep-ph/0606021}}].

\bibitem{Davidson:2007si}
S.~Davidson, G.~Isidori, and S.~Uhlig, {\it {Solving the flavour problem with
  hierarchical fermion wave functions}},  {\em Phys. Lett.} {\bf B663} (2008)
  73--79, [\href{http://xxx.lanl.gov/abs/0711.3376}{{\tt arXiv:0711.3376}}].

\bibitem{Agashe:2009tu}
K.~Agashe, {\it {Relaxing Constraints from Lepton Flavor Violation in 5D
  Flavorful Theories}},  \href{http://xxx.lanl.gov/abs/0902.2400}{{\tt
  arXiv:0902.2400}}.

\bibitem{Iltan:2007sc}
E.~O. Iltan, {\it {The effects of lepton KK modes on the lepton electric dipole
  moments in the Randall Sundrum scenario}},  {\em Eur. Phys. J.} {\bf C54}
  (2008) 583--590, [\href{http://xxx.lanl.gov/abs/0708.3765}{{\tt
  arXiv:0708.3765}}].

\bibitem{Gedalia:2009ws}
O.~Gedalia, G.~Isidori, and G.~Perez, {\it {Combining Direct \& Indirect Kaon
  CP Violation to Constrain the Warped KK Scale}},
  \href{http://xxx.lanl.gov/abs/0905.3264}{{\tt arXiv:0905.3264}}.

\bibitem{Csaki:2008eh}
C.~Csaki, A.~Falkowski, and A.~Weiler, {\it {A Simple Flavor Protection for
  RS}},  \href{http://xxx.lanl.gov/abs/0806.3757}{{\tt arXiv:0806.3757}}.

\bibitem{Cacciapaglia:2007fw}
G.~Cacciapaglia {\em et.~al.}, {\it {A GIM Mechanism from Extra Dimensions}},
  {\em JHEP} {\bf 04} (2008) 006,
  [\href{http://xxx.lanl.gov/abs/0709.1714}{{\tt arXiv:0709.1714}}].

\bibitem{Cheung:2007bu}
C.~Cheung, A.~L. Fitzpatrick, and L.~Randall, {\it {Sequestering CP Violation
  and GIM-Violation with Warped Extra Dimensions}},  {\em JHEP} {\bf 01} (2008)
  069, [\href{http://xxx.lanl.gov/abs/0711.4421}{{\tt arXiv:0711.4421}}].

\bibitem{Santiago:2008vq}
J.~Santiago, {\it {Minimal Flavor Protection: A New Flavor Paradigm in Warped
  Models}},  {\em JHEP} {\bf 12} (2008) 046,
  [\href{http://xxx.lanl.gov/abs/0806.1230}{{\tt arXiv:0806.1230}}].

\bibitem{Csaki:2009bb}
C.~Csaki and D.~Curtin, {\it {A Flavor Protection for Warped Higgsless
  Models}},  \href{http://xxx.lanl.gov/abs/0904.2137}{{\tt arXiv:0904.2137}}.

\bibitem{Csaki:2009wc}
C.~Csaki, G.~Perez, Z.~Surujon, and A.~Weiler, {\it {Flavor Alignment via
  Shining in RS}},  \href{http://xxx.lanl.gov/abs/0907.0474}{{\tt
  arXiv:0907.0474}}.

\bibitem{Arkani-Hamed:2002qy}
N.~Arkani-Hamed, A.~G. Cohen, E.~Katz, and A.~E. Nelson, {\it The littlest
  Higgs},  {\em JHEP} {\bf 07} (2002) 034,
  [\href{http://xxx.lanl.gov/abs/hep-ph/0206021}{{\tt hep-ph/0206021}}].

\bibitem{Cheng:2003ju}
H.-C. Cheng and I.~Low, {\it TeV symmetry and the little hierarchy problem},
  {\em JHEP} {\bf 09} (2003) 051,
  [\href{http://xxx.lanl.gov/abs/hep-ph/0308199}{{\tt hep-ph/0308199}}].


\bibitem{Cheng:2004yc}
H.-C. Cheng and I.~Low, {\it Little hierarchy, little Higgses, and a little
  symmetry},  {\em JHEP} {\bf 08} (2004) 061,
  [\href{http://xxx.lanl.gov/abs/hep-ph/0405243}{{\tt hep-ph/0405243}}].

\bibitem{Buras:2004kq}
A.~J. Buras, A.~Poschenrieder, and S.~Uhlig, {\it {Particle antiparticle
  mixing, epsilon(K) and the unitarity triangle in the littlest Higgs model}},
  {\em Nucl. Phys.} {\bf B716} (2005) 173--198,
  [\href{http://xxx.lanl.gov/abs/hep-ph/0410309}{{\tt hep-ph/0410309}}].

\bibitem{Buras:2006wk}
A.~J. Buras, A.~Poschenrieder, S.~Uhlig, and W.~A. Bardeen, {\it Rare K and B
  decays in the littlest Higgs model without T- parity},  {\em JHEP} {\bf 11}
  (2006) 062, [\href{http://xxx.lanl.gov/abs/hep-ph/0607189}{{\tt
  hep-ph/0607189}}].

\bibitem{Hubisz:2005bd}
J.~Hubisz, S.~J. Lee, and G.~Paz, {\it The flavor of a little Higgs with
  T-parity},  {\em JHEP} {\bf 06} (2006) 041,
  [\href{http://xxx.lanl.gov/abs/hep-ph/0512169}{{\tt hep-ph/0512169}}].

\bibitem{Blanke:2007ww}
M.~Blanke and A.~J. Buras, {\it {A guide to flavour changing neutral currents
  in the littlest Higgs model with T-parity}},  {\em Acta Phys. Polon.} {\bf
  B38} (2007) 2923, [\href{http://xxx.lanl.gov/abs/hep-ph/0703117}{{\tt
  hep-ph/0703117}}].

\bibitem{Buras:2007zt}
A.~J. Buras and C.~Tarantino, {\it {Quark and lepton flavour physics in the
  littlest Higgs model with T-parity}},
  \href{http://xxx.lanl.gov/abs/hep-ph/0702202}{{\tt hep-ph/0702202}}.

\bibitem{Duling:2007sf}
B.~Duling, {\it {Lepton Flavor Violation in the LHT - A Clear Distinction from
  Supersymmetry}},  \href{http://xxx.lanl.gov/abs/0709.4413}{{\tt
  arXiv:0709.4413}}.

\bibitem{Recksiegel:2009vj}
S.~Recksiegel, {\it {Flavour Physics in the Littlest Higgs Model with T-Parity:
  Effects in the $K$, $B_{d,s}$ and $D$ systems}},
  \href{http://xxx.lanl.gov/abs/0908.3117}{{\tt arXiv:0908.3117}}.

\bibitem{Goto:2008fj}
T.~Goto, Y.~Okada, and Y.~Yamamoto, {\it {Ultraviolet divergences of flavor
  changing amplitudes in the littlest Higgs model with T-parity}},  {\em Phys.
  Lett.} {\bf B670} (2009) 378--382,
  [\href{http://xxx.lanl.gov/abs/0809.4753}{{\tt arXiv:0809.4753}}].

\bibitem{delAguila:2008zu}
F.~del Aguila, J.~I. Illana, and M.~D. Jenkins, {\it {Precise limits from
  lepton flavour violating processes on the Littlest Higgs model with
  T-parity}},  \href{http://xxx.lanl.gov/abs/0811.2891}{{\tt arXiv:0811.2891}}.



\bibitem{Artamonov:2008qb}
{\bf E949} Collaboration, A.~V. Artamonov {\em et.~al.}, {\it {New measurement
  of the $K^{+} \to \pi^{+} \nu \bar{\nu}$ branching ratio}},
  \href{http://xxx.lanl.gov/abs/0808.2459}{{\tt arXiv:0808.2459}}.


\bibitem{Blanke:2009pq}
M.~Blanke, {\it {Insights from the Interplay of $K\to\pi\nu\bar\nu$ and
  $\varepsilon_K$ on the New Physics Flavour Structure}},
  \href{http://xxx.lanl.gov/abs/0904.2528}{{\tt arXiv:0904.2528}}.

\bibitem{Ellis:2002fe}
J.~R. Ellis, J.~Hisano, M.~Raidal, and Y.~Shimizu, {\it A new parametrization
  of the seesaw mechanism and applications in supersymmetric models},  {\em
  Phys. Rev.} {\bf D66} (2002) 115013,
  [\href{http://xxx.lanl.gov/abs/hep-ph/0206110}{{\tt hep-ph/0206110}}].

\bibitem{Arganda:2005ji}
E.~Arganda and M.~J. Herrero, {\it Testing supersymmetry with lepton flavor
  violating tau and mu decays},  {\em Phys. Rev.} {\bf D73} (2006) 055003,
  [\href{http://xxx.lanl.gov/abs/hep-ph/0510405}{{\tt hep-ph/0510405}}].

\bibitem{Brignole:2004ah}
A.~Brignole and A.~Rossi, {\it Anatomy and phenomenology of mu tau lepton
  flavour violation in the MSSM},  {\em Nucl. Phys.} {\bf B701} (2004) 3--53,
  [\href{http://xxx.lanl.gov/abs/hep-ph/0404211}{{\tt hep-ph/0404211}}].

\bibitem{Paradisi:2005tk}
P.~Paradisi, {\it Higgs-mediated $\tau\to\mu$ and $\tau\to e$ transitions in II
  Higgs doublet model and supersymmetry},  {\em JHEP} {\bf 02} (2006) 050,
  [\href{http://xxx.lanl.gov/abs/hep-ph/0508054}{{\tt hep-ph/0508054}}].

\bibitem{Paradisi:2006jp}
P.~Paradisi, {\it {Higgs-mediated $e\to \mu$ transitions in II Higgs doublet
  model and supersymmetry}},  {\em JHEP} {\bf 08} (2006) 047,
  [\href{http://xxx.lanl.gov/abs/hep-ph/0601100}{{\tt hep-ph/0601100}}].


\bibitem{Altmannshofer:2009ap}
W.~Altmannshofer, {\it {Probing the MSSM flavor structure with low energy CP
  violation}},  \href{http://xxx.lanl.gov/abs/0909.2837}{{\tt
  arXiv:0909.2837}}.


\bibitem{Agashe:2003rj}
K.~Agashe and C.~D. Carone, {\it {Supersymmetric flavor models and the
  $B\to\phi K_S$ anomaly}},  {\em Phys. Rev.} {\bf D68} (2003) 035017,
  [\href{http://xxx.lanl.gov/abs/hep-ph/0304229}{{\tt hep-ph/0304229}}].

\bibitem{Ross:2004qn}
G.~G. Ross, L.~Velasco-Sevilla, and O.~Vives, {\it {Spontaneous CP violation
  and non-Abelian family symmetry in SUSY}},  {\em Nucl. Phys.} {\bf B692}
  (2004) 50--82, [\href{http://xxx.lanl.gov/abs/hep-ph/0401064}{{\tt
  hep-ph/0401064}}].

\bibitem{Calibbi:2009ja}
L.~Calibbi {\em et.~al.}, {\it {FCNC and CP Violation Observables in a
  SU(3)-flavoured MSSM}},  \href{http://xxx.lanl.gov/abs/0907.4069}{{\tt
  arXiv:0907.4069}}.

\bibitem{Antusch:2007re}
S.~Antusch, S.~F. King, and M.~Malinsky, {\it {Solving the SUSY Flavour and CP
  Problems with $SU(3)$ Family Symmetry}},  {\em JHEP} {\bf 06} (2008) 068,
  [\href{http://xxx.lanl.gov/abs/0708.1282}{{\tt arXiv:0708.1282}}].

\bibitem{Hall:1995es}
L.~J. Hall and H.~Murayama, {\it {A Geometry of the generations}},  {\em Phys.
  Rev. Lett.} {\bf 75} (1995) 3985--3988,
  [\href{http://xxx.lanl.gov/abs/hep-ph/9508296}{{\tt hep-ph/9508296}}].


\bibitem{Buras:2003td}
A.~J. Buras, {\it {Relations between $\Delta M_{s,d}$ and $B_{s,d}\to
  \mu\bar\mu$ in models with minimal flavour violation}},  {\em Phys. Lett.}
  {\bf B566} (2003) 115--119,
  [\href{http://xxx.lanl.gov/abs/hep-ph/0303060}{{\tt hep-ph/0303060}}].

\bibitem{Hurth:2008jc}
T.~Hurth, G.~Isidori, J.~F. Kamenik, and F.~Mescia, {\it {Constraints on New
  Physics in MFV models: a model- independent analysis of $\Delta F=1$
  processes}},  {\em Nucl. Phys.} {\bf B808} (2009) 326--346,
  [\href{http://xxx.lanl.gov/abs/0807.5039}{{\tt arXiv:0807.5039}}].


\bibitem{Lunghi:2008aa}
E.~Lunghi and A.~Soni, {\it {Possible Indications of New Physics in $B_d$
  -mixing and in $\sin(2 \beta)$ Determinations}},  {\em Phys. Lett.} {\bf
  B666} (2008) 162--165, [\href{http://xxx.lanl.gov/abs/0803.4340}{{\tt
  arXiv:0803.4340}}].

\bibitem{Buras:2008nn}
A.~J. Buras and D.~Guadagnoli, {\it {Correlations among new CP violating
  effects in $\Delta F = 2$ observables}},  {\em Phys. Rev.} {\bf D78} (2008)
  033005, [\href{http://xxx.lanl.gov/abs/0805.3887}{{\tt arXiv:0805.3887}}].

\bibitem{Lunghi:2009sm}
E.~Lunghi and A.~Soni, {\it {Hints for the scale of new CP-violating physics
  from $B$-CP anomalies}},  \href{http://xxx.lanl.gov/abs/0903.5059}{{\tt
  arXiv:0903.5059}}.



\bibitem{Buras:2009pj}
A.~J. Buras and D.~Guadagnoli, {\it {On the consistency between the observed
  amount of CP violation in the $K^{-}$ and Bd-systems within minimal flavor
  violation}},  {\em Phys. Rev.} {\bf D79} (2009) 053010,
  [\href{http://xxx.lanl.gov/abs/0901.2056}{{\tt arXiv:0901.2056}}].


\end{thebibliography}
\end{document}